\documentclass{webofc}
\usepackage[varg]{txfonts}   
\begin{document}
\title{SQM2022: Theory Overview in Heavy-Ion Physics}
%
%

\author{\firstname{Joseph} \lastname{Kapusta}\inst{1}\fnsep\thanks{\email{kapusta@umn.edu}} 
}

\institute{School of Physics and Astronomy, University of Minnesota, Minneapolis, Minnesota 55455 USA 
}

\abstract{%
An overview is presented of the many open, interesting questions regarding the behavior of matter at large temperatures and chemical potentials.
}
\maketitle
\section{Introduction}
\label{intro}

The main goal of colliding large nuclei at high energies is to measure the properties of matter under the conditions of extreme temperature and density.  The first heavy ion collision experiments began approximately fifty years ago.  Much has been learned, and much more will be learned, in the coming years.  In a fictional world, one could produce hot dense matter in the accelerator and place it under a microscope for study.   In the real world, theory and experiment must go hand-in-hand to interpet the data and extract physics from it.  

It is not feasible to provide a complete review of the full theoretical effort here.  Rather, the purpose of this overview is to alert the audience to many open physics issues and how they are addressed by the speakers in this conference.

\section{Equation of State}
\label{EOS}

The equation of state of QCD as a function of temperature and density is a fascinating subject with many unanswered questions.

$\bullet$ Is there a critical point in the phase diagram?  Many models suggest so.  Where is the critical curve?

$\bullet$ If a critical point exists it would be in the same universality class as the 3D Ising model and liquid-gas transitions.  How would it be related to a chiral symmetry restoring phase transition, which is in a different universality class, if the up and down quarks were massless?

$\bullet$ To describe the matter created at RHIC beam energies and below, knowledge is required of the equation of state as a function of $T$, $\mu_B$, $\mu_Q$, and $\mu_S$ to conserve energy, baryon number, electric charge, and strangeness.  This is nontrivial when there is critical behavior in the phase diagram.

$\bullet$ Such an equation of state is also needed for modeling neutron star mergers and closely related to the cold dense matter comprising neutron stars.

$\bullet$ Will hard working lattice practioners ever be able to provide the answers on account of the infamous sign problem in the partition function?

\section{Simulations are the Key}
\label{sim}

Analytic models were very useful at the beginning of the field.  Numerical simulations are now essential to make contact between theory and experiment.

$\bullet$ Fluctuations are typically computed in coordinate space but measured experimentally in momentum space.  Dynamical simulations are necessary to compare them.

$\bullet$ Photon and dilepton emission are computed theoretically as functions of the thermodynamic variables.  They must be folded with the space-time evolution of the matter.

$\bullet$ Jet and charm evolution are computed theoretically as functions of the thermodynamic variables.  They must be folded with the space-time evolution of the matter.  Feedback on the matter should be taken into account.

$\bullet$ A sampling of publicly available codes includes

\indent \indent Initial state:  Glauber, LEXUS, IP-Glasma, Trento

\indent \indent Relativistic hydrodynamics:  MUSIC, iEBE-MUSIC, iEBE-VISHNU, \\
\indent \indent SONIC \& superSONIC, SPheRIO, CLVISC, vHLLE, ECHO-QGP

\indent \indent Transport theory: UrQMD, ZPC, AMPT, SMASH

\indent \indent Particlization of hydrodynamics: iSS, Microcanonical Cooper-Frye

\section{Hydrodynamics and Transport Theory}
\label{hydro}

The ``standard model" of relativistic heavy ion collisions includes hydrodynamics, which has proven to be invaluable.  However

$\bullet$ Perfect fluid dynamics is not realistic.  The system is out of equilibrium.  How to define flow velocity?  Should it be via energy flow, baryon flow, electric charge flow, or something else?  This is not an issue for transport models.

$\bullet$ Is the standard model of viscous hydrodynamics coupled with an hadronic afterburner the correct one for lower beam energies? 

$\bullet$ What is the initial condition as it depends on beam energy?

$\bullet$ Transport theory with hadrons and mean fields is undoubtedly a better description at lower beam energies.  It appears to be better able to handle longer nuclear transit times.

$\bullet$  How and when to transition from transport theory to hydrodynamics as a function of beam energy, impact parameter, or even at a fixed impact parameter?  When does the assumption of (approximate) local equilibrium become reasonable?

$\bullet$ Can a theory of spin-hydrodynamics be realized in numerical simulations?

\section{Strangeness}
\label{strange}

Strange quark and anti-quark production is the reason for the origin and existence of this conference series!  

$\bullet$ How does strangeness first appear?  As strange quarks and anti-quarks or as strange mesons and baryons?  How does it depend on beam energy or impact parameter or even as a function of transverse position in a given collision?

$\bullet$ The differences between $\Lambda$ and $\bar{\Lambda}$ and between $K^+$ and $K^-$ are very small at the LHC but become increasingly large as the energy decreases.  What interesting physics can we infer?

$\bullet$ The connection between polarized $\Lambda$ and $\bar{\Lambda}$ baryons, aligned $K^{*0}$ mesons, and vorticity of the matter is extremely interesting.  Current models assume their spins are equilibrized but is that correct?

$\bullet$ How does strangeness evolve?  Is the strange current directly proportional to the baryon current?  If the quarks are polarized, how do they pass that on to the strange mesons and baryons? 

\section{Charm}
\label{charm}

Production of charmed mesons and baryons at RHIC and especially at the LHC is an exciting sub-field with many aspects to it.  These include, but are not limited to the following:

$\bullet$ Charm production in p+p and p+Pb at the LHC is not well reproduced by fragmentation in PYTHIA, which is based on $e^+ e^-$ collisions.  Some physics is missing.

$\bullet$ The charm diffusion coefficient is calculated as a function of $T$ on the lattice and detailed modeling can connect it to experimental data.  What about the chemical potential dependence at lower beam energies?

$\bullet$ Many new exotic charmed baryons and mesons have been detected at the LHC.  Are they tetraquark and pentaquark states or are they molecular states of two hadrons?  

$\bullet$ Do charmed nuclei exist?

$\bullet$ $K^{*0}$ vector mesons in Pb+Pb collisions at the LHC are aligned but $J/\psi$ mesons are not.  What dynamics is involved?

\section{Fluctuations}
\label{fluct}

Relativistic heavy ion collisions can produce thousands, if not tens of thousands, of particles.  Even so, fluctuations are important and can provide valuable information on the dynamics of the system and on the equation of state.  Sources of fluctuations include:

$\bullet$ Initial state fluctuations from randomly chosen locations of nucleons.

$\bullet$ Initial state fluctuations from randomly chosen locations of quarks within nucleons.

$\bullet$ Initial state fluctuations of color gluon field at very high energy.

$\bullet$ Hydrodynamic fluctuations are intimately connected to transport coefficients due to the fluctuation-dissipation theorem.  Inclusion of viscosity and heat conductivity requires them for consistency.

$\bullet$  Hydrodynamic fluctuations in the evolving matter is challenging to implement numerically in 3D.

\section{Correlations}
\label{corr}

Particles can be correlated due to quantum mechanics, conservation laws, and dynamics.  Some intriquing issues are:

$\bullet$ Hanbury-Brown and Twiss intensity interferometry provides information on the space-time evolution of the matter.  Correlations are readily inferred from experiment, but a rigorous derivation for heavy ion collisions is subtle and lacking.

$\bullet$ Charge balance functions can provide information on the charge diffusivity, but extraction from data requires detailed knowledge and modeling of the space-time evolution of the matter.

$\bullet$ How can we understand theoretically the transition in dynamics from small to medium to large collision systems as defined by size of the beam and target or by multiplicity?

$\bullet$ Can Disoriented Chiral Condensates (DCC) be revived at the LHC?

\section{Critical Point and First Order Phase Transition Dynamics}
\label{CP}

Whether or not QCD has a critical point and associated line of first order phase transition in the temperature - chemical potential plane is a major unanswered question.  There are many issues that complicate the matter, including these:

$\bullet$ One may need to be within $10^{-2}$ or $10^{-3}$ of $T_c$ or even less to see the true critical exponents.  The distance is not universal.  Beyond that one typically measures mean field exponents.  This is quite a challenge for heavy ion collisions.

$\bullet$ Crossing a line of first order phase transition in a heavy ion collision ought to be more dramatic than passing near a critical point.

$\bullet$ It is uncertain how hydrodynamic models can deal with a first order phase transition.  Nucleation is the statistical formation of a critical size bubble or droplet of the other phase in the metastable region, but the critical size may be too large for relevance in a heavy ion collision or may not happen fast enough. 

$\bullet$ It is uncertain how hydrodynamic models can deal with a first order phase transition.  Spinodal decomposition occurs when the system crosses into the unstable region where $c_T^2 < 0$.   Subsequent evolution depends on pre-existing fluctuations and inhomogeneities. 

$\bullet$ Transport theories can use mean fields to deal with a two phase system dynamically.  But, fundamentally, does one use hadron or quark and gluon degrees of freedom?  No matter whether a smooth crossover or a critical point, the issue remains.

\section{Neutron Stars and their Mergers}
\label{NS}

Supernovae may reach temperatures of 50 MeV.  Neutron stars may reach energy densities greater than 1 GeV/fm$^3$ with correspondingly high baryon densities.  Mergers of neutron stars, crucially studied with gravitational waves, may equal or exceed both.  This is a great synergy among fields of physics and astrophysics.

$\bullet$ The introduction of new degrees of freedom such as strangeness, in the form of quasi-free quarks or hyperons, softens the equation of state by lowering the pressure at the same energy density.  This makes it challenging to produce neutron stars with two more solar masses.

$\bullet$ A common way to avoid this challenge is to introduce terms in the effective hadronic Lagrangian which makes it energetically unfavorable for strange degrees of freedom to appear.  Can heavy-ion collisions provide quantitative information on these otherwise difficult to constrain terms?

$\bullet$ Interactions between or among nucleons and hyperons are important to constrain the effective Lagrangians.  These interactions can be informed by correlations and HBT of those baryon species.

\section{Conclusion}
\label{conclude}

Asymptotic freedom was discovered almost fifty years ago.  Within two years it was realized that matter should best be described in terms of quark and gluon degrees of freedom at high energy densities.  Since then much analytical and numerical work has been done to elucidate the properties of matter at extreme energy densities.  Theory and experiment have been driving each other to a state of the field that was almost unthinkable in the 1980's.
Keep all this in mind during the conference.  What will the theory summary talk \cite{Berndt} have to say?

I thank Sangyong Jeon, Che-Ming Ko, Volker Koch, Berndt M\"uller, Bj\"orn Schenke and Ramona Vogt for advice on preparing this talk.
This work was supported by the U.S. DOE Grant No. DE-FG02-87ER40328.

%

\begin{thebibliography}{}
%
%
\bibitem{Berndt}
B. M\"uller, theory summary in this conference.
\end{thebibliography}
%
%

\end{document}